\documentclass[twocolumn,showpacs,preprintnumbers,amsmath,amssymb,superscriptaddress,aps,prl]{revtex4}

\usepackage{graphicx}
\usepackage{dcolumn}
\usepackage{bm}

\begin{document}

\preprint{APS/123-QED}

\title{Compact atomic gravimeter based on a pulsed and accelerated optical lattice}

\author{Manuel~Andia}
\affiliation{Laboratoire Kastler Brossel, Ecole Normale Sup\'erieure, Universit\'e Pierre et Marie Curie, CNRS, 4 place Jussieu, 75252 Paris Cedex 05, France}
\author{Rapha\"el~Jannin}
\affiliation{Laboratoire Kastler Brossel, Ecole Normale Sup\'erieure, Universit\'e Pierre et Marie Curie, CNRS, 4 place Jussieu, 75252 Paris Cedex 05, France}
\author{Fran\c cois~Nez}
\affiliation{Laboratoire Kastler Brossel, Ecole Normale Sup\'erieure, Universit\'e Pierre et Marie Curie, CNRS, 4 place Jussieu, 75252 Paris Cedex 05, France}
\author{Fran\c cois~Biraben}
\affiliation{Laboratoire Kastler Brossel, Ecole Normale Sup\'erieure, Universit\'e Pierre et Marie Curie, CNRS, 4 place Jussieu, 75252 Paris Cedex 05, France}
\author{Sa\"\i da~Guellati-Kh\'elifa}
\affiliation{Laboratoire Kastler Brossel, Ecole Normale Sup\'erieure, Universit\'e Pierre et Marie Curie, CNRS, 4 place Jussieu, 75252 Paris Cedex 05, France}
\affiliation{Conservatoire National des Arts et M\'etiers,
292 rue Saint Martin, 75141 Paris Cedex 03, France}
\email{guellati@spectro.jussieu.fr}
\author{Pierre~Clad\'e}
\affiliation{Laboratoire Kastler Brossel, Ecole Normale Sup\'erieure, Universit\'e Pierre et Marie Curie, CNRS, 4 place Jussieu, 75252 Paris Cedex 05, France}

\date{\today}

\begin{abstract}
We present a new scheme of compact atomic gravimeter based on atom interferometry. Atoms are maintained against gravity using a sequence of coherent accelerations performed  by the Bloch oscillations technique. We demonstrate a sensitivity of 4.8$\times 10^{-8}$ with an integration time of 4 min. Combining this method with an atomic elevator allows to measure the local gravity at different positions in the vacuum chamber. This method can be of relevance to improve the measurement of the Newtonian gravitational constant $G$. 
\end{abstract}

\pacs{37.25.+k,37.10.Jk,06.30.Gv,06.20.Jr}
\maketitle

Atom interferometry has proven to be a reliable method to realize robust inertial sensors~\cite{Kasevich1991,Peters,Gustavson2000a,LeGouet2008,Gauguet2009,Schmidt2011,Stockton2011a,Tackmann2012}. The performance of these devices rivals state-of-the-art sensors based on other  methods. The record sensitivity of 8$\times 10^{-9}g$  at 1~s is achieved for the measurement of the gravity acceleration $g$ \cite{Muller2008,Louchet-Chauvet2011}, allowing precise determination of the gravity gradient and the Newtonian gravitational constant $G$\cite{Fixler2007,Lamporesi2008}. In such atom interferometers, the inertial phase shift scales quadratically with the interrogation time. An accurate measurement of the gravity acceleration requires a long time of free-fall and should be limited by the size of the vacuum cell in which the measurement takes place. 
The best performance was obtained using cold atoms launched along a parabolic trajectory of about 1 m \cite{Peters}. This constraint limits the development of compact and transportable atomic gravimeters necessary to high-precision geophysical on-site measurements. Furthermore, the value of the gravity is averaged over a large height.

Atom gradiometers stemming from free-fall gravimeters are constrained to shorter measurement times. Some of these gradiometers are used to measure the Newtonian gravity constant \cite{Fixler2007,Lamporesi2008}. As related in these references, the largest contribution  to the error budget comes from the uncertainty on the atomic cloud position within the fountain and the initial launch velocity of the atoms. To improve the measurement of the $G$ constant, there is a need for a conceptually different gravimeter capable of locally measuring gravity  at a well-controlled position. 

Several methods were proposed to implement a compact atomic gravimeter. They consist in levitating the atoms by means of the laser light \cite{Impens2006, Hughes2009, DeSaint-Vincent2010, Poli2011, Clade2005, Charriere2012, Pelle2013}. Currently three kinds of experiments based on atoms trapped in a vertical optical lattice are in progress~\cite{Clade2005,Poli2011, Charriere2012,Pelle2013}: using atoms confined in an amplitude-modulated vertical optical lattice~\cite{Poli2011}, using atom interferometry involving a coherent superposition between different Wannier-Stark atomic states in a 1-D optical lattice \cite{Pelle2013} and combining a Ramsey-Bord\'e interferometer with Bloch oscillations in a quasi-stationary vertical standing wave~\cite{Clade2005, Charriere2012}. In the last experiment atoms interact with the optical lattice in the middle of the interferometer sequence during  about 100 ms. They reach the best short-term sensitivity of  $3.5\times$10$^{-6} g$ at 1s. This sensitivity is limited by a contrast decay of the interference fringes due to the decoherence induced by the inhomogeneity of the lattice laser beams~\cite{Charriere2012}. 

\begin{figure}
\begin{center}
\includegraphics[width = \linewidth]{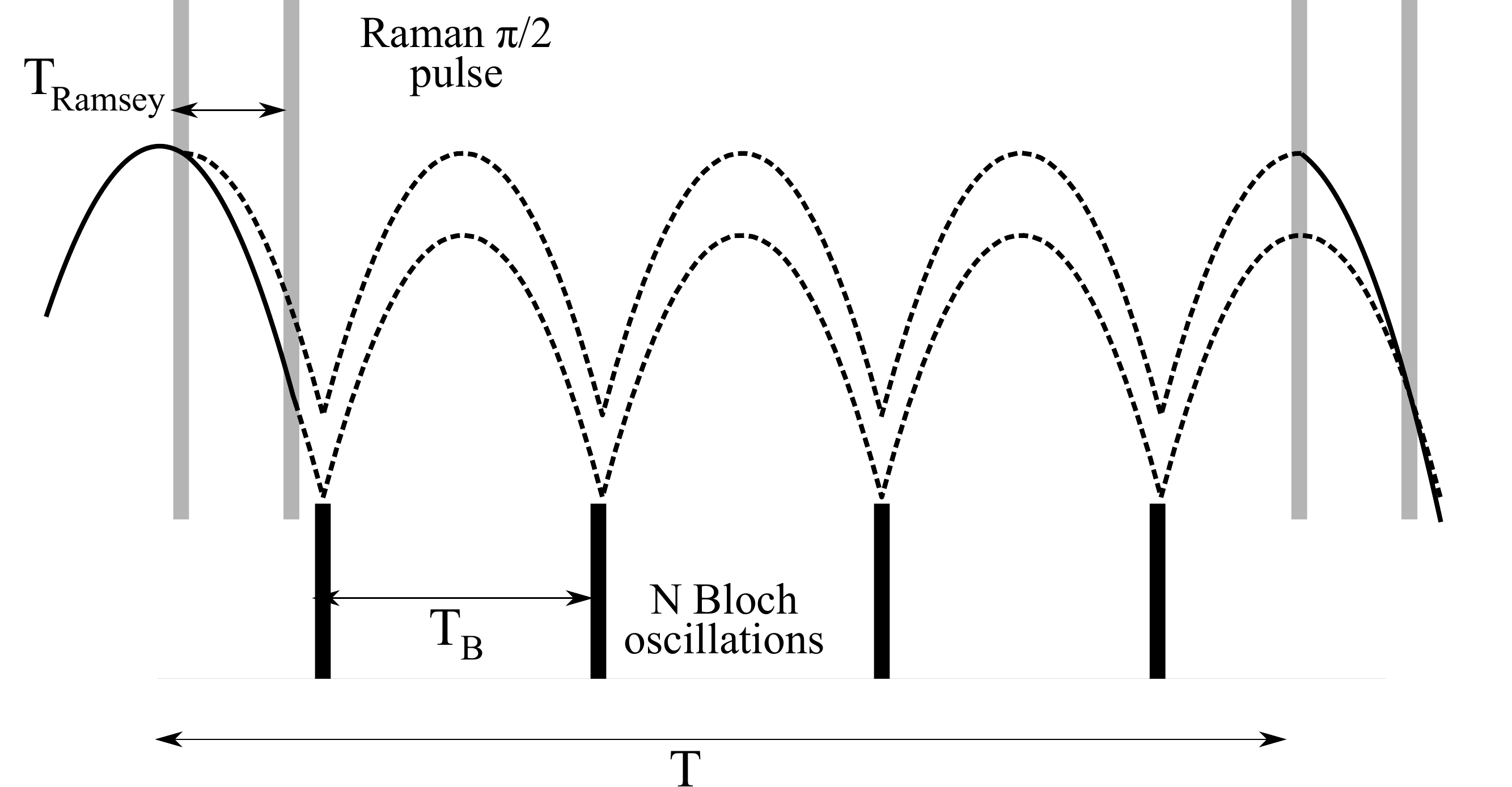}
\end{center}
\caption{Principle of the method. We use laser-cooled $^{87}$Rb atoms in $F=2$. They are first prepared with an initial velocity close to zero (trajectory in solid curve), we apply a first pair of $\pi/2$ pulses to select atoms in $F=1$ (dashed curve). A laser beam resonant with the D2 line pushes away atoms remaining in state $F=2$ (those trajectories are not shown in this figure). Then we pulse and accelerate the optical lattice to hold the atoms in $F=1$  against gravity, in the middle of the atom interferometer sequence. }
\label{figure1}
\end{figure}

In this paper we demonstrate a new method to measure precisely the local acceleration of gravity. The principle is illustrated in figure~\ref{figure1}. It is based on a Ramsey-Bord\'e atom interferometer realized by two pairs of $\pi/2$ Raman pulses. In order to compensate the fall of atoms between the pulses, we use a sequence of brief and strong accelerations. The acceleration is based on the method of Bloch oscillations in an accelerated optical lattice. Because the lattice is pulsed we have less decoherence compared to previously described methods where the force is applied continuously.

We obtain a preliminary  measurement of the local gravity acceleration with a short-term sensitivity of  $7.4\times$10$^{-7}$ at 1 s. The atoms are maintained within a 4.6 mm falling distance during about 230 ms. Using similar acceleration pulses, it is possible to control precisely the position and the initial velocity of the atoms before the beginning of the atom interferometer.

\medskip

The principle of the acceleration process consists in transferring to the atoms many photon recoils by the means of Bloch Oscillations (BO)~\cite{Peik,BenDahan,Wilkinson1996}. This is done by a succession of Raman transitions in which the atom begins and ends in the same energy level, so that its internal state  is unchanged while its velocity has increased by 2$v_\mathrm{r}$ after each transition ($v_\mathrm{r}=\hbar k/m$ is the recoil velocity of the atom of mass $m$ when it absorbs
a photon of momentum $\hbar k$). BO are produced in a one-dimensional
vertical optical lattice which is accelerated by linearly
sweeping the relative frequencies of the two counter-propagating
laser beams. This leads to a succession of rapid
adiabatic passages between momentum states differing by $2\hbar k$. The Bloch oscillation technique offers a remarkable ability to coherently and efficiently transfer photon momenta~\cite{Battesti04}. 

In \cite{Charriere2012}, the configuration of Bloch oscillations in a vertical standing wave has been deeply investigated. The authors have observed a drop of the contrast when the number of Bloch oscillations is increased, limiting this number to 75 (corresponding to 100 ms). 
This result differs from what was observed in accelerated lattices where a contrast of 30\% is observed for 600 BO performed in 4 ms~\cite{Cadoret2008}. The contrast decay is due to the speckle pattern, which induces a random force on the atoms. Indeed this force is proportional to the depth $U_0$ of the optical lattice. The phase imprinted by the speckle is proportional to the lattice depth times its duration.  However 
the critical acceleration which sets the efficiency of Bloch oscillation scales as $U_0^2$ (adiabaticity criterion), in the weak binding limit \cite{Peik,BenDahan}. In our experiment the depth of the lattice is 10 times as high  as in the experiment using a standing wave, whereas the light is switched on for a duration 50 times shorter. The decoherence due to the inhomogeneities of the Bloch laser beams is therefore lower using a sequence of brief and strong accelerations from BO.

\medskip
\begin{figure}
\begin{center}
\includegraphics[width = \linewidth]{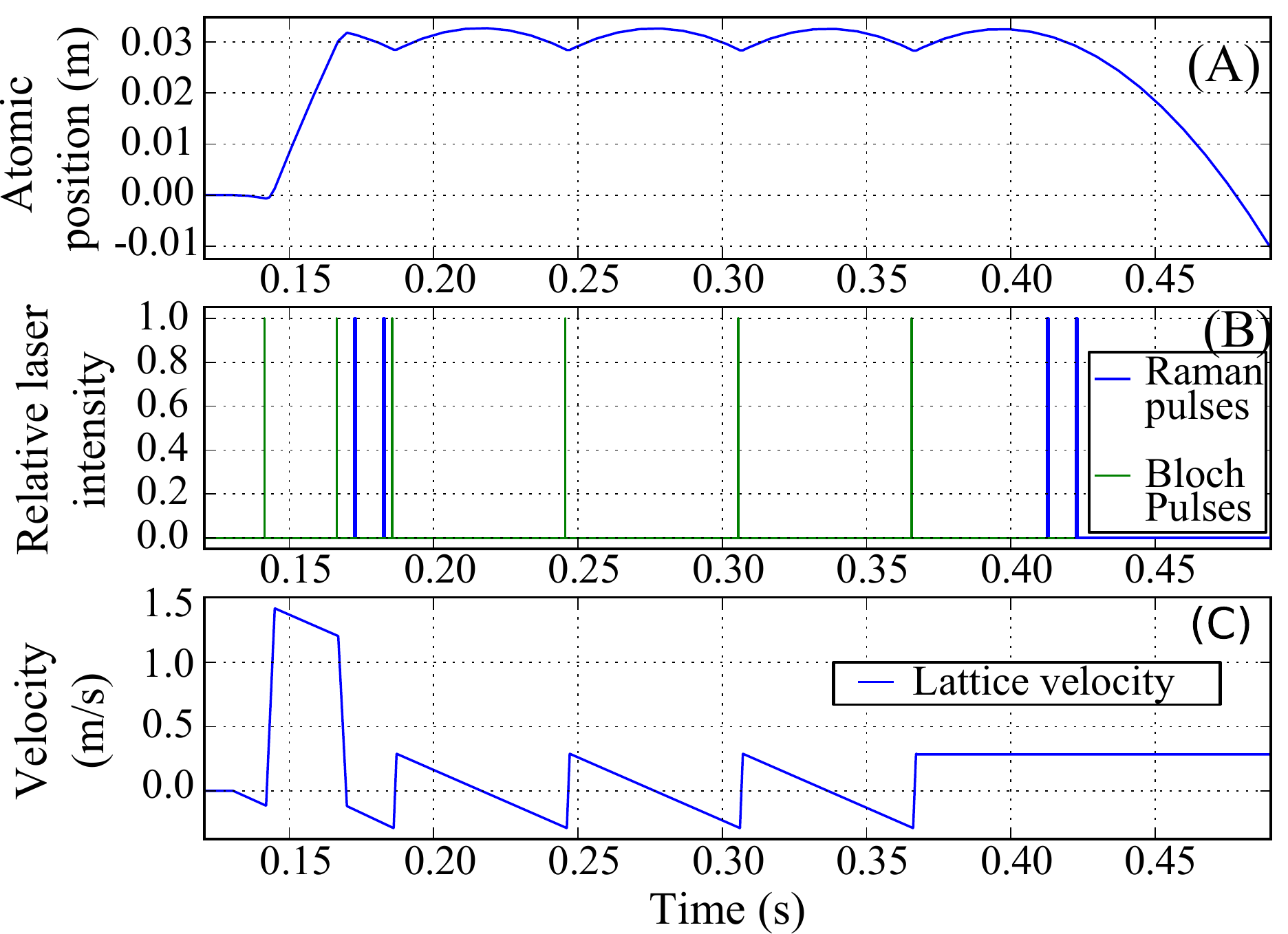}
\end{center}
\caption{(Color online) (A): trajectory of the atoms during the measurement procedure. The atoms are first  held at a given position using an atomic elevator. We apply the first pair of $\pi/2$ Raman pulses, then atoms are maintained against gravity during 230 ms by periodically transferring them  100 photon momenta.\\(B): timing sequence of the Raman and Bloch beams, the two first Bloch pulses are used to perform the atomic elevator.\\(C): velocity of the optical lattice versus time.}
\label{figure2}
\end{figure}

We use a Ramsey-Bord\'e atom interferometer realized by two pairs of $\pi/2$ laser pulses. Each light pulse induces a Doppler-sensitive Raman transition which couples the hyperfine levels $F=1$ and $F=2$ of the $^5S_{1/2}$ ground state of $^{87}$Rb. 
The first pair of $\pi/2$ pulses transfers the atoms from the $F$=2
hyperfine level to $F$=1 and selects the initial velocity distribution. 
The second pair measures  the final velocity of the atoms by transferring resonant atoms from $F$=1 to $F$=2. Note that after the first pair of $\pi/2$ pulses we apply a laser pulse resonant with the D2 line in order to push away atoms remaining in the state $F=2$.  Atoms in $F$=1 perform  $M$ series of $N$ Bloch oscillations between the two pairs of $\pi/2$ pulses : after the first pair we let the atoms fall during $T_B$, then they are shone with the accelerated optical lattice. They acquire a velocity of 2 $N\times v_\mathrm{r}$ in the upward direction. The delay $T_B$ is chosen in such  a way that the gravity acceleration is perfectly compensated by the coherent acceleration due to Bloch oscillations, \textit{i.e} $T_B=2Nv_\mathrm{r}/g$. This process is repeated periodically to maintain the atoms against gravity. Figure~\ref{figure1} depicts only the trajectories of atoms in $F=1$ after the first pair of $\pi/2$ pulses.

To maintain the atoms within a short falling distance, it is necessary to prepare them with an initial velocity close to zero. For this purpose, the atom inteferometer sequence is preceded by the \textit{atomic elevator} sequence\cite{Cadoret2008} (see figure~\ref{figure2}). It consists in two sets $N_1$ and $N_2$ of BO: atoms are first accelerated in a given direction using $N_1$ BO. When they reach the chosen position they are stopped by using $N_2$ BO in the opposite direction. The final position of the atoms and their velocity are precisely determined by the numbers $N_1$ and $N_2$ and the spacing 
time between the two Bloch pulses. This method allows us to displace the atoms, without losses, at different positions in the vacuum chamber, before starting the measurement of the local gravity.

The experiment is realized in a titanium UHV-chamber connected to a glass cell by a differential pumping tube. It is shielded from residual magnetic fields by two layers of $\mu$ metal. 
The two-dimensional magneto optical trap (2D-MOT) produces a
slow $^{87}Rb$ atomic beam (about 10$^9$ atoms/s at a velocity of
20 m/s) which loads during 250 ms a three-dimensional
magneto optical trap. Then a $\sigma^+$-$\sigma^-$ molasses generates
a cloud of about 2 $\times$ 10$^8$ atoms in the $F =2$ hyperfine
level, with a 1.7 mm radius and at a temperature of 4 $\mu$K.

The Bloch beams originate from a 3.8 W Ti:sapphire laser pumped by an 18 W@532 nm laser (Verdi G18-Coherent). The output laser beam is split into two paths, each of which passes through an acousto-optic modulator (AOM) to adjust the frequency offset and amplitude before being injected into a polarization-maintaining fiber. The depth of the generated optical lattice is 21$E_r$ (for $^{87}Rb$ the recoil energy is $E_r \simeq 3.77$ kHz$\times h$, where $h$ is the Planck constant.) for an effective power of  220 mW seen by the atoms. The Bloch lasers are blue-detuned by 30 GHz from the rubidium D2 line. Under these conditions, the Landau-Zener tunneling and the scattering rate are negligible. The Bloch pulses are shaped by controlling the radio-frequency signal driving the acousto-optic modulators (AOM). The power lattice is raised on in 500 $\mu s$, in order to adiabatically load  the atoms in the first Brillouin zone. The number of Bloch oscillations is determined by fixing the frequency chirp of the RF signal used to drive the AOMs. The Raman and Bloch beams are collimated to a 1/$e^2$ diameter of 11 mm at the output of the  polarization-maintaining fibers used to guide light toward  the vacuum chamber (for details of the optical setup see~\cite{Bouchendira2011,Cadoret2008}). 
The timing sequence of Raman and Bloch pulses during  the experiment is presented on figure~\ref{figure2}-B. The delay $T_\mathrm{R}$ between two $\pi/2$ pulses is 10 ms and the duration of each pulse $\tau$ equals 1.3 ms. Figure~\ref{figure2}-C shows the velocity of the optical lattice as a function of time.

\medskip

\begin{figure}
\begin{center}
\includegraphics[width = \linewidth]{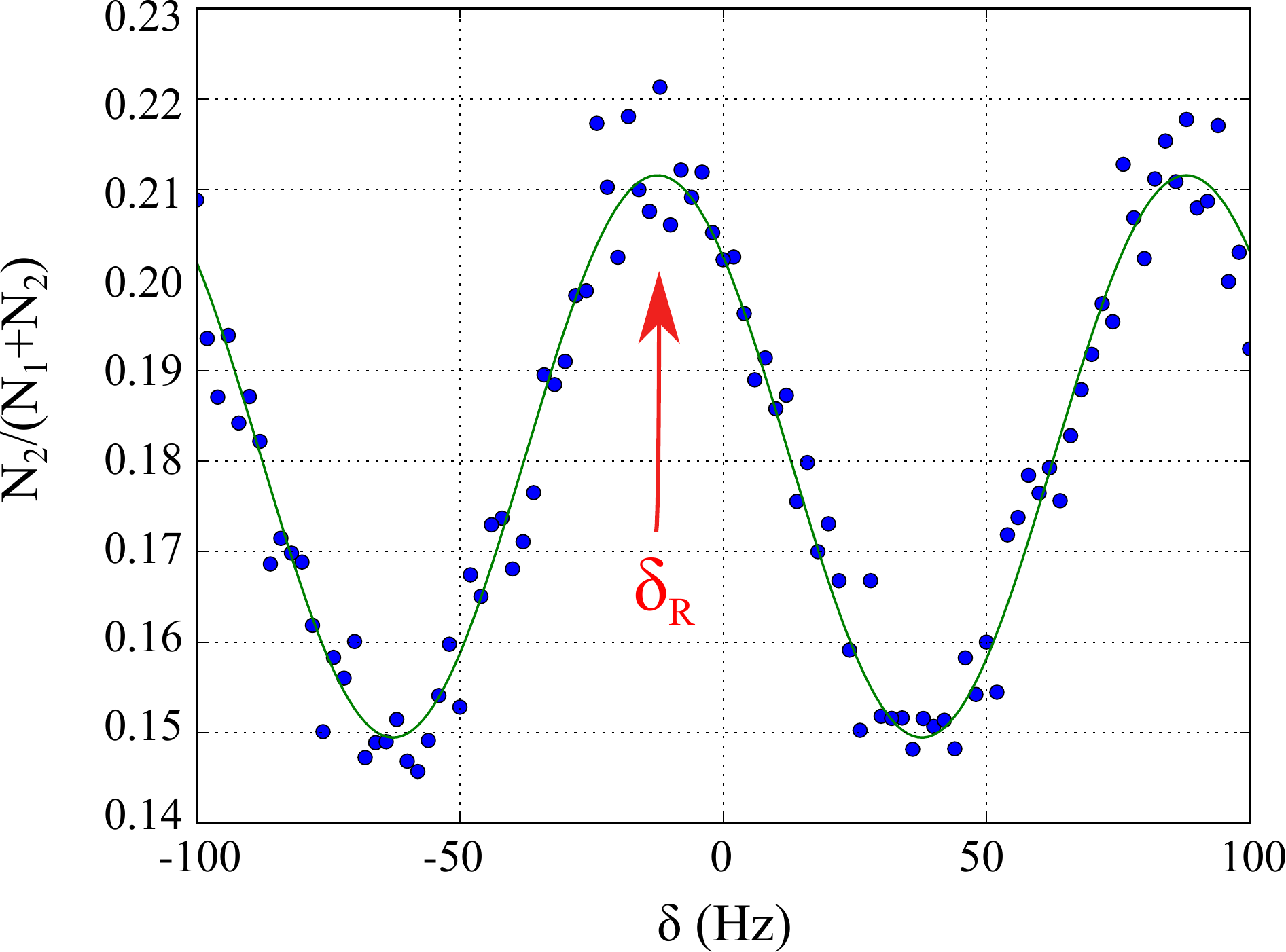}
\end{center}
\caption{(Color online) Typical spectrum used to deduce a value of $g$. The data points show the quantity $N_2/(N_1+N_2)$ where $N_1$ and $N_2$ are the populations in the $F=1$ and $F=2$ levels, measured for different  values of $\delta$, the change of the Raman frequency between the two pairs of $\pi/2$ pulses. The solid line is a least-squares fit to the data by a $\cos((\delta - \delta_R)\times T_R)$ function. The position $\delta_R$ of the central fringe is determined from this fit.}
\label{figure3}
\end{figure}

A way to determine the value of $g$ would consist in scanning the value of  the delay $T$ between the two pairs of $\pi/2$ pulses, keeping the relative Raman frequency $\delta$ unchanged. If the delay $T$ equals exactly  2$N M \times v_\mathrm{r}/g$, the BO will
compensate exactly the acceleration due to gravity and the two paths of the interferometer will be in phase.
In the experiment we set the value of $T$ close but no exactly equal to the
expected value. The remaining velocity $\delta v$ will induce a shift
$\delta_R = (k_1+k_2)\times\delta v$ of the position of the fringes when we scan
$\delta$. A typical interference fringe obtained with $N$=50, $M$=4 and $T$=227 ms is shown in figure~\ref{figure3}. The data points show the relative population in $F=2$ 
measured by the second pair of $\pi/2$ pulses as the Raman frequency $\delta$ is
scanned. The fringe shift $\delta_R$ is determined by a least-squares fit of the experimental data by a $\cos((\delta - \delta_R)\times T_R)$ function. In practice, we operate the interferometer close to the central fringe, which is previously located by using different values of $T_R$.

The frequency shift $\delta_R$ is related to the local gravity acceleration by
\begin{equation}
g=\frac{1}{T}\times\left( \frac{2NM\hbar k_B}{m_{\mathrm{Rb}}}- \frac{\delta_\mathrm{R}}{(k_1+k_2)}\right) 
\label{eq1}
\end{equation}
where $k_\mathrm{B}$ is the Bloch wave vector, $k_1$ and $k_2$ are the wave vectors of the two Raman beams. The ratio $h/m_{\mathrm{Rb}}$ between the Planck constant and the rubidium atomic mass $m_\mathrm{Rb}$ is measured by our group with a relative uncertainty of 1.2$\times 10^{-9}$~\cite{Bouchendira2011}.  In equation~\ref{eq1}, the term $\delta_\mathrm{R}/(k_1+k_2)$ is small and appears as a correction to the reference value $g_{\mathrm{ref}}=2NMv_\mathrm{r}/T$. As a consequence, to achieve a precise determination of $g$, the precision on the value of the Raman wave-vectors ($k_1$ and $k_2$) is less critical than on the value of the Bloch wave-vector. 

\begin{figure}
\begin{center}
\includegraphics[width = \linewidth]{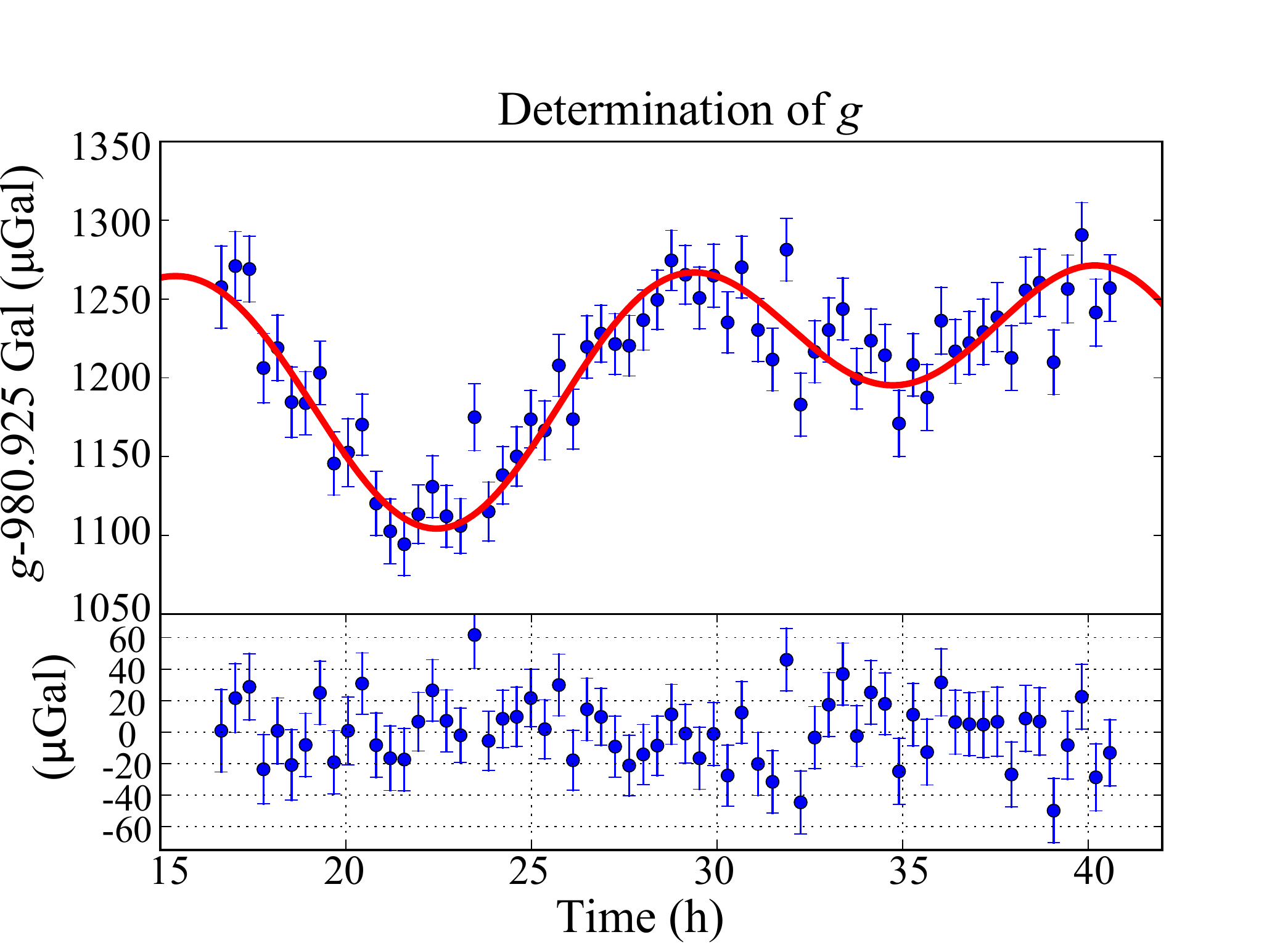}
\end{center}
\caption{(Color online) Gravity data taken over one day fitted by the earth tide model. Each data point is deduced from the average over 6 measurements. The lower curve shows residuals of the fit (1$\mu$Gal=10$^{-8}$ m/s$^2$).}
\label{figure4}
\end{figure}

As the fall distance is small, the systematic errors due to the gradients of residual magnetic fields and light fields are negligible. In practice, to cancel the parasitic effect due to the temporal fluctuations of these fields, we record two spectra exchanging the direction of the Raman beams. We achieve a relative statistical uncertainty of 4.8$\times 10^{-8}$ on the value of $g$ for an integration time of 4 min. Figure~\ref{figure4} shows a temporal behavior of the gravity measured over 25 hours. These data fit well with a local solid Earth tide model \cite{Wenzel1993,Tamura1987}, the temporal fluctuations of local gravity are dominated by tidal forces.

\begin{figure}
\begin{center}
\includegraphics[width = \linewidth]{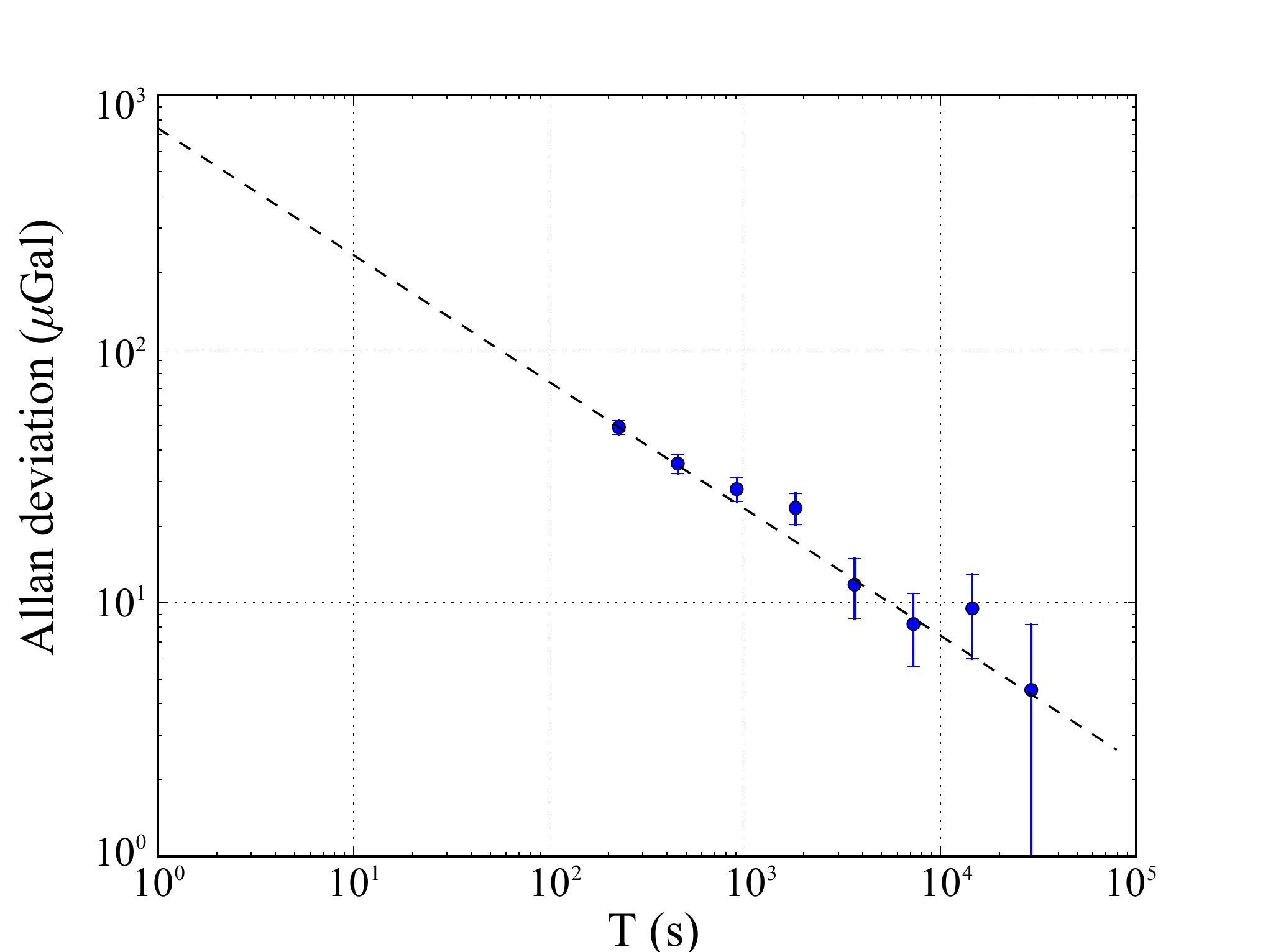}
\end{center}
\caption{(Color online) Allan standard deviation of the local gravity, for a delay $T$ = 227 ms and a number of BO $N$=50 (1$\mu$Gal=10$^{-8}$ m/s$^2$).}
\label{figure5}
\end{figure}

The sensitivity of the gravimeters is characterized by the Allan standard deviation of the $g$ measurement. Figure~\ref{figure5} shows the  Allan standard deviation of the set of 388 determinations of $g$; it scales as $t^{-1/2}$ (where $t$ is measurement time). The short-term sensitivity, extrapolated to 1 s according to the white noise behavior, is 7.4$ \times 10^{-7}g$.

Thanks to the "atomic elevator" sequence which precedes the measurement of $g$, it is possible to change the position at which gravity is measured. Because we use BO in the elevator, the acceleration of atoms is well known and it is possible to precisely calculate the displacement of the cloud, using the numbers $N_1$ and $N_2$ of BO and the duration of the elevator. Our setup can therefore be used to precisely measure the gravity gradient.
However, the sensitivity on $g$ is not high enough to map the gravity gradient, whose order of magnitude is $\sim$ 310 $\mu$Gal/m (1$\mu$Gal=10$^{-8}$ m/s$^2$). In this work we have focused on the sensitivity of this new method. The systematic errors which affect the value of $g$ are similar to the ones identified in usual gravimeters (Gouy phase, wavefront aberrations, Coriolis force, level shifts,...) \cite{Peters,Clade2006,Louchet-Chauvet2011}. They should be investigated and presented in an upcoming article. 

\medskip
 
In this paper we have demonstrated a new method to locally measure the gravitational acceleration. It is based on a Ramsey-Bord\'e inteferometer and a sequence of Bloch oscillations. We obtain a preliminary sensitivity of 7.4$\times$10$^{-7}$ at 1s. This sensitivity can be improved using a colder atomic source (100 nK) and by reducing the vibrations to achieve a delay $T_\mathrm{R}$ of about 50 ms. With these experimental parameters we should achieve a sensitivity comparable to the state of the art.
The key feature of our method lies in the decoherence induced by the fluctuations of the optical lattice  which is substantially reduced compared to similar methods. 

We notice that in our atomic interferometer the frequencies of the Raman beams are the same during  the first and the second pairs of $\pi/2$ pulses. We can imagine implementing a succession of atom interferometers where the last pair of $\pi/2$ pulses of each interferometer is the first pair of the next one. In such a configuration the phase noise between successive interferometers will be correlated and the Allan variance should decrease as $1/n$ (where $n$ is the number of measurements).

We have elaborated  a timing sequence, which allows us to move the atoms toward different positions in the vacuum chamber before performing the local gravity measurement. Potential applications of this method to improve the measurement of the Newtonian gravitational constant $G$ should be investigated. In the experiment reported in \cite{Fixler2007}, the constant $G$ is determined using a gradiometer which  measures the differential acceleration of two samples of laser-cooled
atoms. The change in the gravitational field along one dimension is measured when a test mass is displaced along a distance of 27.940 cm. The main systematic errors come from the uncertainties on the position and the initial velocity of the atoms. More recently the experiment of Tino et al.\cite{Lamporesi2008} improved  the uncertainty on $G$ by one order of magnitude. The main contribution to the systematic error on
the G measurement derives from the initial velocity of the atomic cloud and the positioning accuracy of the source masses. The authors claimed that the latter should be reduced  by about 1 order of magnitude by using a laser tracker. We propose to investigate the method described in this paper to reduce the systematic error due to atomic cloud parameters.

This experiment is supported in part by IFRAF (Institut
Francilien de Recherches sur les Atomes Froids), the \textit{Agence Nationale pour la Recherche}, FISCOM Project-(ANR-06-BLAN-0192), the\textit{ Direction G\'en\'erale de l'armement (DGA)} and "\textit{Emergence}"  program of university Pierre et Marie Curie. 

We thank Alexandre Bresson, Nassim Zahzam, Malo Cadoret and Yannick Bidel for providing us a tide model using the internal ONERA software based on the Tamura/Wenzel model. 
\bibliography{apssamp}

\end{document}